\documentclass[twocolumn,preprintnumbers,prd,superscriptaddress,nofootinbib,floatfix,showpacs,showkeys]{revtex4-1}

\usepackage{amsmath}
\usepackage{amsfonts}
\usepackage{amssymb}
\usepackage{graphicx}
\usepackage{color}
\usepackage{hyperref}
\usepackage{booktabs}
\usepackage{hyperref}
\usepackage{braket}
\usepackage{mathtools}
\usepackage[rightcaption]{sidecap}
\usepackage{subfigure}
\usepackage{soul}
\usepackage{comment}

\usepackage{enumerate}

\definecolor{vividviolet}{rgb}{0.62, 0.0, 1.0}
\definecolor{amaranth}{rgb}{0.9, 0.17, 0.31}
\definecolor{palatinateblue}{rgb}{0.15, 0.23, 0.89}
\definecolor{brightpink}{rgb}{1.0, 0.0, 0.5}

\hypersetup{ linktoc=all,
    colorlinks, linkcolor={blue},
    citecolor={brightpink}, urlcolor={palatinateblue}
}

\begin{document}
\title{Black hole thermodynamics from logotropic fluids}

\author{Salvatore Capozziello}
\email{capozziello@na.infn.it}
\affiliation{Dipartimento di Fisica ``E.~Pacini", Universit\`a di Napoli ``Federico II", Via Cinthia 9, 80126 Napoli, Italy.}
\affiliation{Scuola Superiore Meridionale, Largo San Marcellino 10, 80138 Napoli, Italy.}
\affiliation{Istituto Nazionale di Fisica Nucleare (INFN), Sezione di Napoli, 80126 Napoli, Italy.}

\author{Rocco D'Agostino}
\email{rocco.dagostino@unina.it}
\affiliation{Scuola Superiore Meridionale, Largo San Marcellino 10, 80138 Napoli, Italy.}
\affiliation{Istituto Nazionale di Fisica Nucleare (INFN), Sezione di Napoli, 80126 Napoli, Italy.}

\author{Alessio Lapponi}
\email{alessio.lapponi-ssm@unina.it}
\affiliation{Scuola Superiore Meridionale, Largo San Marcellino 10, 80138 Napoli, Italy}
\affiliation{Istituto Nazionale di Fisica Nucleare (INFN), Sezione di Napoli, 80126 Napoli, Italy}

\author{Orlando Luongo}
\email{orlando.luongo@unicam.it}
\affiliation{Divisione di Fisica, Universit\`a di Camerino, Via Madonna delle Carceri 9, 62032 Camerino, Italy.}
\affiliation{Dipartimento di Matematica, Universit\`a di Pisa, Largo B. Pontecorvo 5, 56127 Pisa, Italy.}
\affiliation{Institute of Experimental and Theoretical Physics, Al-Farabi Kazakh National University, 050040 Almaty, Kazakhstan.}

\begin{abstract}
We show that the Einstein field equations with a negative cosmological constant can admit black hole solutions whose thermodynamics coincides with that of logotropic fluids, recently investigated to heal some cosmological and astrophysical issues. For this purpose, we adopt the  Anton-Schmidt equation of state, which represents a generalized version of logotropic fluids. We thus propose a general treatment to obtain an asymptotic anti-de Sitter metric, reproducing the thermodynamic properties of both Anton-Schmidt and logotropic fluids. Hence, we explore how to construct suitable spacetime functions, invoking an event horizon and fulfilling the null, weak, strong and dominant energy conditions. We further relax the strong energy condition to search for possible additional solutions. Finally, we discuss the optical properties related to a specific class of metrics and show how to construct an effective refractive index depending on the spacetime functions and the thermodynamic quantities of the fluid under study. We also explore possible departures with respect to the case without the fluid.  
\end{abstract}


\maketitle

\section{Introduction}

Over the last years, studies on the black hole (BH) entropy have revealed  close connections between the  thermodynamic properties and the event horizon of a BH \cite{Hawking:1971vc,Bardeen:1973gs,Bekenstein:1973ur}. Among all different kinds of BHs provided with different geometries and thermodynamic features, Schwarzschild BH represents the simplest case, where part of the radiation is absorbed by the BH mass \cite{Hawking:1975vcx,Wald:1999vt}. Other interesting BH solutions include Reissner-Norstr\"om BH, whose thermodynamics is similar to that of regular BH \cite{Myung:2007av, Ghaffarnejad:2013wpa,Good:2020qsy}, Ho\v{r}ava-Lifshitz BH \cite{Cai:2009qs,JahaniPoshteh:2021clv}, characterized by rich thermodynamic properties  and so forth.

Moreover, asymptotic BH solutions to the de Sitter space can be obtained from the Einstein equations with a positive cosmological constant $(\Lambda>0)$ \cite{Gibbons:1977mu,Liu:2016urf}. In such a case, it has been shown that the surface gravity of the BH horizon would determine the temperature of particles emitted from the BH \cite{Pappas:2017kam}. The same, however, happens even for the cosmological event horizon, so that a thermal equilibrium may occur only if the two surfaces coincide \cite{Davies:1988dma,Chatterjee:2020gfc,Donnay:2019zif}. Further, BHs that are asymptotic to the anti-de Sitter (AdS) space can be found as solutions to the Einstein field equations with $\Lambda<0$ \cite{Anabalon:2012tu}. As in the case of the asymptotically flat space, the entropy and temperature of AdS BH are equal to 1/4 of the event horizon area, whereas, differently from the flat space case, such objects admit, at a given temperature, a stable equilibrium with radiation, and a positive specific heat \cite{Hawking:1982dh}. Throughout recent years, the physics of asymptotically AdS BHs has gained a renewed interest due to the AdS/CFT duality \cite{Maldacena:1997re,Lunin:2001jy,Hubeny:2014bla}. In this context, particular attention was given  to the study of thermal field theories living on the AdS boundary and, from the bulk perspective, to the several phase transitions that these types of BH exhibit. 

Furthermore, in treating the cosmological constant as the thermodynamic pressure, it has been shown that the thermodynamics of a charged/rotating AdS BH exactly coincides with that of Van der Waals's fluid \cite{Rajagopal:2014ewa,Roy:2021ucl,Capozziello:2004ej}. Subsequently, an asymptotically AdS solution to the Einstein equations was obtained by matching the BH thermodynamic parameters with those of a particular class of polytropic gas \cite{Setare:2015xaa}. Then, an additional AdS BH solution was found in the thermodynamic framework of modified Chaplygin gas \cite{Debnath:2019mzs,Kamenshchik:2001cp}.

Motivated by those findings, in the present study, we focus on the class of logotropic models, whose  thermodynamic features permit to heal astrophysical issues related to dark matter distribution in galaxies, and unify the cosmological dark sector \cite{Chavanis:2015paa,Chavanis:2015eka,Chavanis:2016pcp,Benaoum:2021pqf}. As a prominent result, these models can be generalized to the well-known Anton-Schmidt fluid \cite{ANTON1997449,ASfluid}. These scenarios have been recently proposed in the cosmological context as a unified dark energy model \cite{Capozziello:2017buj,Capozziello:2018mds,Boshkayev:2019qcx,Odintsov:2018obx}\footnote{For alternative approaches to dark energy, see also \cite{DAgostino:2019wko,DAgostino:2021vvv,Capozziello:2019cav,Capozziello:2022wgl,Capozziello:2022rac,Bajardi:2022tzn,DAgostino:2022tdk}.}. In this respect, the Anton-Schmidt fluid has been also studied in the Tolman-Oppenheimer-Volkov formalism \cite{Tolman:1939jz,Oppenheimer:1939ne} to obtain analytical solutions for a static and spherically symmetric BH \cite{Capozziello:2020szy}. In particular, from the relation between the Anton-Schmidt free parameters and the BH mass, one can find spacetime solutions describing Schwarzschild-de Sitter BH and naked singularities. Thus, it appears natural to investigate the thermodynamic consequences to check whether the inclusion of a logotropic and/or the Anton-Schmidt equation of state (EoS) may lead to reasonable results in the BH description.

In this paper, we search for a BH solution to the Einstein field equations, whose corresponding thermodynamics coincides with that of logotropic models. Starting from the  Anton-Schmidt EoS,  we propose a general treatment to obtain an asymptotic Schwarzschild-AdS metric, which reproduces the thermodynamic properties of the involved fluids, i.e. the pressure and the density. In particular, we motivate this choice since in a homogeneous and isotropic universe, those quantities appear crucial in order to write the energy-momentum tensor, as it will be clarified  later in the text.
Thus, to determine the most suitable metric functions, we present a general method involving any possible density term contribution. Moreover, in order to have a physical BH, we invoke the existence of an event horizon and investigate under which circumstances the  energy conditions may hold. We also explore the possibility of violating the strong energy condition, in order to find additional physical properties. We then discuss the physical consequence of this recipe in view of the free constants emerging from the integration procedure. With the aim of distinguishing among different thermodynamic BHs, we consider the optical properties of our solutions and show how to construct an effective refractive index, following the standard procedure adopted for static and spherically symmetric spacetimes. In particular, we show that the net dependence of the refractive index on the underlying spacetime can lead to different outcomes.  The refractive index increases significantly under the choice of particular constant values, whereas the asymptotic regime is investigated in terms of density, showing the limit to Schwarzchild-AdS. Hence, we explore possible departures with respect to the case without the logotropic fluid, corresponding to a pure Schwarzschild-AdS case.

The paper is organized as follows. After this introduction, in Sec.~\ref{sec:A-S BH} we introduce the Anton-Schmidt EoS and its limit to logotropic models. There we postulate the metric ansatz for a static, spherically symmetric metric that is consistent with an asymptotic AdS spacetime. We thus analyze the thermodynamic properties of the Anton-Schmidt BH in terms of its mass, temperature and entropy. In Sec.~\ref{sec:energy conditions}, we constrain BH solutions requiring the presence of an event horizon and checking the validity of the energy conditions. In particular, we discuss how the violation of the strong energy condition may lead to a metric solution containing a factor that can be associated with a refractive index. Finally, in Sec.~\ref{sec:conclusions}, we summarize our findings and draw the conclusions of our work. In this study, we use Planck units $c=\hbar=G=1$.

\section{Logotropic black holes}
\label{sec:A-S BH}

Let us start by considering the Einstein field equations with the cosmological constant in the form
\begin{equation}
G_{\mu\nu}+\Lambda g_{\mu\nu}=8\pi T_{\mu\nu}\,,
\label{eq:EFE}
\end{equation}
where $G_{\mu\nu}\equiv R_{\mu\nu}-\frac{1}{2}Rg_{\mu\nu}$ is the Einstein tensor, $g_{\mu\nu}$ is the metric tensor, and $T_{\mu\nu}$ is the stress-energy tensor of the source fluid.
According to recent studies 
\cite{Kubiznak:2012wp,Gunasekaran:2012dq}, in the extended phase space one can interpret $\Lambda$ as a thermodynamic pressure, namely\footnote{Alternatively, it is possible to work out the same recipe adopting the conjugate variable of pressure, namely the volume \cite{Kubiznak:2012wp}.}
\begin{equation}
P=-\dfrac{\Lambda}{8\pi}=\frac{3}{8\pi l^2}\,,
\label{eq:thermo P}
\end{equation}
where $l$ is the AdS curvature constant.
Our aim is to construct an asymptotic AdS BH whose thermodynamics matches that of the Anton-Schmidt fluid with pressure given by 
\begin{equation}\label{ASfluid}
    P=A\left(\frac{\rho}{\rho_\ast}\right)^{-n}\ln\left(\frac{\rho}{\rho_\ast}\right),
\end{equation}
where the density $\rho$ is normalized to a reference density $\rho_\ast$, while $A>0$ and $n\neq -1$ are constants. 

This class of fluid has been introduced in \cite{ASfluid} for crystalline solids, where the Anton-Schmidt EoS gives the empirical expression of crystalline solid’s pressure under isotropic deformation. Afterwards, in the field of cosmology, see e.g. \cite{Capozziello:2017buj}, it has been argued that, in analogy with solid state physics, the pressure naturally changes its sign, showing how the cosmic speed-up naturally emerges as the universe volume changes under the action of cosmic expansion. To account for this mechanism, one can assume the $n$ parameter to depend upon the  Gr\"uneisen index, $\gamma_G$, i.e., $n=n(\gamma_G)$, related to the specific heat at constant volume and to the bulk modulus. This semi-empirical relation provides a temperature dependence of the free parameter $n$ that can be tested experimentally. We here consider fixing the index $n$ to a constant, namely without assuming the temperature dependence through the  Gr\"uneisen index. In fact, this allows one to investigate a given epoch of the universe dynamics that may correspond to our time, where the temperature effects are negligible. Hence, we shall model our BH configuration through the pressure and the density only, which clearly represent the main ingredients of the energy-momentum tensor at late times. In so doing, our black hole configuration shows a cosmological constant contribution to density and pressure that matches the Anton-Schmidt fluid given by Eq.~\eqref{ASfluid}.

Our strategy is to start from Eq.~\eqref{ASfluid}, which generalizes the logotropic models with $n=0$. In so doing, we recover the logotropic thermodynamics as a limiting case of the Anton-Schmidt fluid. 
The logotropic thermodynamics is of utmost importance, especially in the framework of dark matter configuration. Indeed, it is possible to show that pressureless dark matter leads to cuspy density profiles, disfavoured by observations that, instead, suggest a constant density core. If the dark matter halo shows a polytropic EoS, that describes both dark matter
halos and the cosmological evolution, then the logotropic solution appears as the most natural one. Again, we thus require that our main thermodynamical properties to investigate are pressure and density as above reported. In addition to what we discussed above, for the sake of completeness, generalized versions of logotropic models have been also investigated \cite{Capozziello:2018mds,Benaoum:2021pqf,Chavanis:2022vzi} and criticized, see e.g. \cite{Boshkayev:2021uvk}, but lie beyond the purposes of this work.

Bearing in mind the above considerations, we shall consider the static and spherically symmetric line element
\begin{equation}
ds^2=f(r,\rho)dt^2-\frac{dr^2}{f(r,\rho)}-r^2d\Omega^2\,,
\label{eq:metric}
\end{equation}
where 
\begin{equation}\label{Cgeneralization}
    f(r,\rho)=-\frac{2M}{r}+\frac{r^2}{l^2}-h(r,\rho)\,.
\end{equation}
Here, $M$ is the BH mass, and $h(r,\rho)$ is an unknown auxiliary function to determine. 
Clearly, in the case of $h(r,\rho)\rightarrow0$, the postulated metric represents an asymptotic AdS spacetime. In this scheme, the term $h(r,\rho)$ represents a correction when it is considered different from zero. 
Using Eq.~\eqref{eq:thermo P}, we can rewrite Eq.~\eqref{Cgeneralization} as
\begin{equation}\label{metricAnsatzgeneral}
    f(r,\rho)=-\frac{2M}{r}+\frac{8}{3}\pi r^2 P-h(r,\rho)\,.
\end{equation}
In what follows, we seek a class of metrics such that the BH thermodynamics predicted by the latter coincide exactly with the Anton-Schmidt EoS, containing the logotropic models in the limit $n=0$. 
The cosmological pressure $P$, affecting the BH thermodynamics, can be therefore associated with the pressure of the fluid $P(\rho)$. In this way, the function $h(r,\rho)$ accounts for the metric correction that occurs by considering an EoS different from the case of a pure cosmological constant\footnote{Although degenerating with a pure cosmological constant, the case of dark fluid \cite{Luongo:2018lgy,DAgostino:2022fcx,Belfiglio:2022egm} has not been explored here.}, namely $P=-\rho$.

\subsection{Black hole thermodynamics}\label{BHth}

To find the most suitable form of $h(r,\rho)$, we start from the standard BH entropy in terms of the horizon radius $r_h$ and area $A$ as \cite{Bekenstein:1973ur,Bekenstein:1974ax,Bekenstein:1975tw}
\begin{equation}
S=\frac{A}{4}=\pi r_h^2\,.
\label{eq:entropy}
\end{equation}
We can thus relate the BH thermodynamic properties to the parameters of the Anton-Schmidt and logotropic fluids. In particular, the BH mass could be obtained from the definition of horizon radius, namely $f(r_h,\rho)=0$:
\begin{equation}\label{mass}
    M=\frac{4}{3}\pi r_h^3 P-\frac{r_h}{2}h(r_h,\rho)\,.
\end{equation}
Since the EoS of the cosmological constant is $\rho+P=0$, the enthalpy associated with $\Lambda$ is vanishing. For a BH with volume $V$, the total energy within $V$ is $E=M-PV$, and then $M=E+PV$. Hence, it is natural to associate the mass of the BH with its enthalpy $H$, such that $M=H(S,P)$ \cite{Kastor:2009wy,Dolan:2010ha}.

One can thus use the standard thermodynamics relations to calculate the volume and temperature of the BH by exploiting Eqs.~\eqref{eq:entropy} and \eqref{mass}:
\begin{align}
    & V=\left(\dfrac{\partial H}{\partial P}\right)_S=\frac{4\pi r_h^3}{3}-\frac{r_h}{2}\frac{\partial h(r_h,\rho)}{\partial \rho}\left(\frac{\partial P}{\partial \rho}\right)^{-1}\,,  \label{volume} \\
    & T= \left(\dfrac{\partial H}{\partial S}\right)_P=2r_h P-\frac{1}{4\pi r_h}\frac{\partial (rh(r,\rho))}{\partial r}\bigg|_{r_h}\,. \label{temperature}
\end{align}

From the first law of thermodynamics, $dE=TdS-PdV$, and assuming the following integrability condition
\begin{equation}
\frac{\partial^2 S}{\partial T \partial V}=\frac{\partial^2 S}{\partial V \partial T}\,,
\end{equation}
one finds
\begin{equation}\label{int1stlaw}
    S=\left(\dfrac{\rho+P}{T}\right)V\,.
\end{equation}
Therefore, plugging Eqs.~\eqref{ASfluid}, \eqref{eq:entropy}, \eqref{volume} and \eqref{temperature} into Eq.~\eqref{int1stlaw} and considering the solution for a generic $r\geq r_h$, we obtain
\begin{equation}\label{diffeq}
    8\pi r^2(P-2\rho) P_\rho+6(\rho+P) h_\rho-3(rh)' P_\rho=0\,,
\end{equation}
where the subscript $\rho$ and the prime  denote the partial derivatives with respect to the density and radial coordinate, respectively. 
Starting from the theoretical setup presented in \cite{Rajagopal:2014ewa}, we seek a solution of Eq.~\eqref{diffeq} for a  generic $r$ by implementing a general method that makes use of  combinations of linearly independent functions of the density. In particular, a similar approach has been employed in  \cite{Setare:2015xaa, Debnath:2019mzs}, but imposing \emph{a priori} the functional expressions for $R_i(\rho)$. In our treatment, we relax this hypothesis  as illustrated in more detail in appendix \ref{appendixA}. We thus write
\begin{equation}\label{addLinComb}
    h(r,\rho)=\sum_{i}X_i(r)R_i(\rho)\,,
\end{equation}
where the coefficients $X_i$ depend on the radial coordinate, $r$. In this way,  Eq.~\eqref{diffeq} takes the formal expression
\begin{equation}\label{eqs}
    \sum_j\xi_j(r)F_j(\rho)=0\,,
\end{equation}
where $j$ labels the linearly independent density functions, $F_j(\rho)$, accounting for the information on the fluid EoS, and  the relative weights, $\xi(r)$. 
In this way, we obtain a system of equations, whose solution provides the coefficients of $h(r,\rho)$, i.e. $X_i(r)$. 
We report all the calculations in the case of the Anton-Schmidt fluid in appendices \ref{appendixA} and \ref{appendixB}. The final result leads to
\begin{equation}
\label{addtermfinal}
    h(r,\rho)=\frac{c_1}{r}+\frac{8}{3}\pi r^2P(\rho)+c_2r^{\frac{2}{B_2}-1}(a(\rho))^{1/B_2}\,.
\end{equation}   
where $c_1$, $c_2$ and $B_2$ are free parameters, whereas  
\begin{equation}
    a(\rho)\equiv \exp\left(\int \frac{ P_\rho }{P+\rho} \,d\rho\right).
    \label{eq:a}
\end{equation}
Clearly, the above solution leads to a $f(r,\rho)$ that, once plugged into Eq.~\eqref{Cgeneralization}, implies a solution to the vacuum field equations plus a constant term that mimics the constant arising from the AdS assumption made above. 

In the following, we analyze the solutions resulting from Eq.~\eqref{addtermfinal} and the relative energy conditions.

\section{Metric solutions and energy conditions}
\label{sec:energy conditions}

Inserting $h(r,\rho)$ from Eq.~\eqref{addtermfinal} into Eq.~\eqref{metricAnsatzgeneral}, we find the metric solution
\begin{equation}\label{generalmetric}
    f(r,\rho)=-\frac{2M+c_1}{r}-c_2r^{\frac{2}{B_2}-1}a(\rho)^{1/B_2}\,.
\end{equation}
Here, $c_1$ has the only effect to modify the value of the BH mass without affecting the underlying physics\footnote{Even if we do not consider changing the mass sign, repulsive gravity effects are possible in arbitrary spacetimes, see e.g. \cite{Luongo:2014qoa,Luongo:2015zaa,Giambo:2020jjo}.}, so that we can safely set  $c_1=0$. Also, to lighten the notation, we redefine $B_2=\beta$.

In order to obtain an event horizon, we impose the condition $f(r_h,\rho)=0$, namely
\begin{equation}\label{radiusAS}
    r_h=\left(\frac{2M}{-c_2}\right)^{\frac{\beta}{2}}\frac{1}{\sqrt{a(\rho)}}\,,
\end{equation}
requiring the condition $c_2<0$. Therefore, for simplicity, hereafter we set $c_2=-1$. 

A further condition is
$f(r,\rho)\rightarrow 0$ as approaching the horizon from outside, i.e. for $r\rightarrow r_h$. 
We thus require $f(r,\rho)>0$ in the range $r>r_h$, namely  $f(r_h+\epsilon,\rho)>0$ for very small $\epsilon$: 
\begin{align}
&   0<-\frac{2M}{r_h+\epsilon}+(r_h+\epsilon)^{\frac{2}{\beta}-1}a(\rho)^{1/\beta}  \nonumber  \\   
 & \quad   \sim f(r_h,\rho)+\epsilon\left[\frac{2M}{r_h^2}+r_h^{\frac{2}{\beta}-2}\left(\frac{2}{\beta}-1\right)a(\rho)^{1/\beta}\right]. 
\end{align}
Since $f(r_h,\rho)=0$ by definition, the last inequality becomes
\begin{equation}\label{condpositivehorizon}
  \frac{2M}{r_h^2}+r_h^{\frac{2}{\beta}-2}\left(\frac{2}{\beta}-1\right)(a(\rho))^{1/\beta} >0\,.
\end{equation}
Now we shall insert the expression for the horizon radius \eqref{radiusAS} into \eqref{condpositivehorizon}, obtaining
\begin{equation}
\begin{split}
    &(2M)^{1-\beta}a(\rho)+(2M)^{1-\beta}\left(\frac{2}{\beta}-1\right)a(\rho)>0\,,
\end{split}
\end{equation}
which implies $\beta>0$ in order to have a BH.

It is worth mentioning that the horizon existence is mathematically plausible, although, physically,  the corresponding compact object can also be different from a BH\footnote{In other words, any BH is provided with a horizon, but not all objects exhibiting horizons are BHs.}. To better clarify this point, one may consider the Hartle-Thorne metric \cite{Papapetrou:1966zz}, whose well-known horizon  describes massive and compact stars \cite{Stephani:2003tm}. However, the size of these objects is usually larger than the size of the event horizon, so the latter has not a precise meaning. Moreover, the Hartle-Thorne metric possesses its internal counterpart, which is meaningless for the event horizon, albeit mathematically one can always calculate it in analogy to the Sun and Earth. For the aforementioned reasons, we intend to clarify that our BH solution is that of a thermodynamic BH, i.e. a BH-like counterpart exhibiting the same thermodynamics of logotropic models.
Nevertheless, this does not limit the validity of our solution. In principle, we can  apply our outcomes to some precise cases, e.g. compact objects, quite different from genuine BH configurations. In fact, our recipe may be extended to other cases that may require, in astrophysics, an event horizon to determine.

\subsection{Matching the energy conditions}

In order to guarantee the energy conditions to hold and to investigate their effects in the framework of our solution, we shall now  compute the stress-energy tensor induced by Eq.~\eqref{eq:metric}. It is convenient to adopt the tetrad formalism, in which the stress-energy tensor can be written as
\begin{equation}
    T^{\mu\nu}=\varrho e_0^\mu e_0^\nu+\sum_{i=1}^3p_ie_i^\mu e_i^\nu\,.
    \label{energy-momentum}
\end{equation}
The latter becomes the source of the Einstein field equations that might be imposed in order to obtain a BH solution. Indeed, we consider the  static and spherically symmetric line element prompted in Eq.~\eqref{eq:metric} and we impose that this metric satisfies the Einstein equations by virtue of the additional condition found in Eq.~\eqref{generalmetric}. Specifically, since we are interested in having a de Sitter-like contribution, under the form of cosmological constant, as in Eq.~\eqref{eq:thermo P}, we obtain that the Einstein field equations give
\begin{align}
    \varrho&=-p_1=\frac{1-f-rf'}{8\pi r^2}+P\,, \label{varrhogeneral} \\
    p_2&=p_3=\frac{rf''+2f'}{16\pi r}-P\,. \label{transpressuregeneral}
\end{align}
These results agree with the outcomes given in \cite{Rajagopal:2014ewa,Setare:2015xaa,Debnath:2019mzs}, where $p_1$  the longitudinal pressure, whereas $p_2$ and $p_3$ are the transversal pressures that will be denoted by $p_{tr}$. As above stressed, the term $P$, entering  Eqs.~\eqref{varrhogeneral} and \eqref{transpressuregeneral}, is requested to provide the cosmological constant  presence in the Einstein field equations \eqref{eq:EFE}.
Therefore, our solution resembles a BH, since it guarantees that the Einstein equations are solved. However, we stress that the obtained objects are more similar to BH-mimickers, whose existence has been recently raised  \cite{mima}. These configurations have been investigated even without external thermodynamic sources, but assuming only dark energy fields (see e.g. \cite{kis,vis}). Clearly, future efforts are needed to disclose the nature of our solutions in view of the arising BH astronomy.
It is worth  remarking that $\rho$ is the density of the Anton-Schmidt fluid, while $\varrho$ is the density arising in the spacetime whose thermodynamics is emulated by such a fluid.
After simple manipulations, Eqs.~\eqref{varrhogeneral} and \eqref{transpressuregeneral} read
\begin{align}
\varrho&=P-\frac{a(\rho)^{1/\beta}}{4\pi \beta}r^{\frac{2}{\beta}-3}+\frac{1}{8\pi r^2}\,, \\
p_{tr}&=\left(\frac{1}{\beta}-\frac{1}{2}\right)\frac{a(\rho)^{1/\beta}}{4\pi \beta}r^{\frac{2}{\beta}-3}-P\,.
\end{align}

It is worth noting that the energy density and the pressure of the Anton-Schmidt/logotropic fluid
have different notation, as one can see in Eqs.~\eqref{varrhogeneral} and \eqref{transpressuregeneral}. Indeed, we built up a thermodynamic analog of the BH horizon, including the cosmological constant. 
The above energy density and pressure, in general, depend on the radial coordinate. The assumption to consider the analog fluid density $\rho$ to be independent of $r$ is only for simplification purposes and may be generalized. In any case, Eqs.~\eqref{energy-momentum}, \eqref{varrhogeneral} and \eqref{transpressuregeneral} show that our choice leads to localized solutions of the Einstein field equations \cite{Bayin86}.

We can now restrict the possible solutions of Eq.~\eqref{generalmetric}, investigating when the energy conditions are satisfied. Specifically, we have:
\begin{itemize}
    \item the \textit{null energy condition} (NEC): $\varrho+p_i\ge0$ for $i=1,2,3$. In our case $p_2=p_3$ and $p_1=-\varrho$. Thus, the NEC becomes $\varrho+p_{tr}\ge 0$;
    \item the \textit{weak energy condition} (WEC): $\varrho\ge0$. 
    \item the \textit{strong energy condition} (SEC): $\varrho+\sum_{i}p_i\ge0$, which, in our case, becomes $p_{tr}\ge0$;
    \item the \textit{dominant energy condition} (DEC): $\rho\ge|p_i|$ that becomes $\rho\ge|p_{tr}|$.
\end{itemize}We can prove that there is only one metric function $f(r,\rho)$ satisfying all of them. In particular, the WEC reads
    \begin{equation}\label{WEC}
    P-\frac{a(\rho)^{1/\beta}}{4\pi \beta}r^{\frac{2}{\beta}-3}+\frac{1}{8\pi r^2}\ge0\,, 
\end{equation}
while the SEC becomes
\begin{equation}\label{SEC}
    \left(\frac{1}{\beta}-\frac{1}{2}\right)\frac{a(\rho)^{1/\beta}}{4\pi \beta}r^{\frac{2}{\beta}-3}-P\ge0\,.
\end{equation}
Moreover, the NEC condition becomes 
\begin{equation}\label{NEC}
    \varrho+p_{tr}=\frac{1}{8\pi r^2}+\left(\frac{1}{\beta}-\frac{3}{2}\right)\frac{a(\rho)^{1/\beta}}{4\pi \beta}r^{\frac{2}{\beta}-3}\ge0\,.
\end{equation}
The inequality for the DEC depends on the sign of the longitudinal and transverse pressures. In our case, if the SEC is satisfied, the DEC becomes
\begin{equation}
    2P+\frac{1}{8\pi r^2}-\left(\frac{1}{\beta}+\frac{1}{2}\right)\frac{a(\rho)^{1/\beta}}{4\pi \beta}r^{\frac{2}{\beta}-3}\ge0\,.
\end{equation}

Let us check when the WEC and the SEC are satisfied considering four different ranges of values for $\beta$.
\begin{enumerate}[i.]
\item $0<\beta<2/3$. In this case, the term $\frac{a(\rho)^{1/\beta}}{4\pi \beta}r^{\frac{2}{\beta}-3}$ in the WEC dominates when $r$ is high, since $\frac{2}{\beta}-3>2$. However, this term is also negative, therefore the WEC is not satisfied for high radii.
\item $2/3<\beta\le2$. One can easily prove that, in this case, the WEC is satisfied when
\begin{equation}\label{WECinequality}
    P(\rho)r^2-\frac{a(\rho)^{1/\beta}}{4\pi \beta}r^{\frac{2}{\beta}-1}+\frac{1}{8\pi}\ge0\,.
\end{equation}
Since $0\le\frac{2}{\beta}-1<2$, the WEC is satisfied both in the limits $r\rightarrow0$ and $r\rightarrow\infty$. However, it may happen that the WEC is not satisfied in a certain finite range of $r$. Nevertheless, ignoring this possibility, the SEC is certainly not satisfied for high radii, since the dominant term becomes $-P$.
\item $\beta>2$. For small radii, we have a negative term in the WEC dominating over the positive ones. Thus, the WEC is not satisfied for small radii.
\item $\beta=2/3$. The WEC and the SEC read, respectively, 
 \begin{align}
     P-\frac{3}{8\pi}a(\rho)^{3/2}+\frac{1}{8\pi r^2}\ge0\,, \\
     \frac{3}{8\pi}a(\rho)^{3/2}-P\ge0\,.
 \end{align}
 Since they must be satisfied for all the radii, we conclude that  
 \begin{equation}\label{energyconditiontot}
     P(\rho)=\frac{3}{8\pi}a(\rho)^{3/2}.
 \end{equation}
  Thus, $\varrho=-p_1=\frac{1}{8\pi r^2}$ and $p_{tr}=0$, which implies that also the NEC and the DEC are satisfied.
 Moreover, we note that the asymptotic AdS spacetime is exactly recovered in this case. Indeed, inserting Eq.~\eqref{energyconditiontot} into Eq.~\eqref{generalmetric}, we obtain
\begin{equation}
    f(r,\rho)=-\frac{2M}{r}+\frac{8}{3}\pi P(\rho)r^2\,.
\end{equation}
\end{enumerate}
 
 Now, we shall study the restrictions on the parameters of the Anton-Schmidt fluid emulating the situation $\beta=2/3$ just described. From Eq.~\eqref{energyconditiontot}, since $a(\rho)$ is positive, it is clear that the pressure must be positive. By applying the logarithm on both sides of Eq.~\eqref{energyconditiontot}, we obtain
\begin{equation}\label{u7}
    \frac{3}{2}\int\frac{ P_\rho}{P+\rho}d\rho=\ln\left(\frac{8\pi}{3}P\right),
 \end{equation}
where we have used the definition in  Eq.~\eqref{eq:a}. 
Then, it is possible to solve Eq.~\eqref{u7} by simply computing the derivative with respect to $\rho$. Thus, we immediately find
\begin{itemize}
    \item[--] a physical solution, namely $P=const$,
    \item[--] a unphysical solution, namely $P=2\rho$.
\end{itemize}
Even though the first case appears appealing, it just represents the widely-studied trivial cosmological constant case. It appears clear that, in order to have it from Eq.~\eqref{ASfluid}, one needs $\rho=const$, having both the pressure and density to be perfectly constant. On the other hand, the second case  would imply a sound speed faster than light. 
Consequently, in the case of the Anton-Schmidt and logotropic models, an interesting scenario would arise from the violation of at least one energy condition. 

In view of the aforementioned considerations, in what follows we analyze the case that corresponds to relaxing the SEC.

\subsection{Relaxing the strong energy condition}

As previously stated, it appears interesting to relax the SEC. Previously, we computed  possible values of $h(r,\rho)$ in Eq.~\eqref{addtermfinal}, in order for the WEC and the SEC to hold. In so doing, we proved that two solutions for $P$ satisfy both of them, implying automatically that the NEC and the DEC hold as well. 
Since the only plausible outcome is the one with constant pressure, it is of utmost importance to investigate a wider range of possible solutions for Eq.~\eqref{generalmetric}.

We limit ourselves to those  ranges where  WEC is satisfied\footnote{As SEC is not satisfied in this ranges, $p_{tr}<0$ and consequently, the DEC coincides with the NEC, Eq.~\eqref{NEC}. So, the NEC is necessary and sufficient to prove the DEC.}, having $2/3<\beta\le2$.
Then, the NEC (as the DEC) becomes
\begin{equation}\label{DECineq}
    \left(\frac{1}{\beta}-\frac{3}{2}\right)\frac{a(\rho)^{1/\beta}}{4\pi \beta}r^{\frac{2}{\beta}-3}+\frac{1}{8\pi r^2}\ge0\,.
\end{equation}
The above condition is never satisfied when $2/3<\beta<2$, since the first term is negative in this range and it dominates over the second. The only possibilities are then $\beta=2/3$ or $\beta=2$. 
In the first case, we have exactly the same situation studied before satisfying the SEC, i.e. Eq.~\eqref{energyconditiontot}. 
Instead, in the case $\beta=2$, we find that the WEC, the NEC and the SEC are all satisfied when  $\frac{a(\rho)^{1/\beta}}{4\pi \beta}\le\frac{1}{8\pi}$, implying
\begin{equation}\label{energycondpartial}
   \exp\left(\frac{1}{2}\int\frac{P_\rho }{\rho+P}\, d\rho\right)\le1\,.
\end{equation}
Applying the logarithm to both sides, we obtain
\begin{equation}
    \int\frac{P_\rho}{\rho+P}\,d\rho\le0 \,.
\end{equation}
which is satisfied as long as the integral is upper-bounded. 
In the case of the Anton-Schmidt and pure logotropic fluids, we respectively have
\begin{equation}\label{derivativesign}
   \frac{P_\rho}{P+\rho}=\frac{\frac{A}{\rho_\ast}\left(\frac{\rho}{\rho_\ast}\right)^{-n-1}\left[1-n\ln\left(\frac{\rho}{\rho_\ast}\right)\right]}{\rho+A\left(\frac{\rho}{\rho_\ast}\right)^{-n}\ln\left(\frac{\rho}{\rho_\ast}\right)} \,,
\end{equation}
and
\begin{equation}\label{derivativesignBIS}
   \frac{P_\rho}{P+\rho}=\frac{A}{\rho\left[\rho+A\ln\left(\frac{\rho}{\rho_\ast}\right)\right]} \,.
\end{equation}

The condition $P=\frac{|\Lambda|}{8\pi}$ ensures the positivity of $P(\rho)$ and, thus, $\rho >\rho_\ast$. The sign of the right side of Eq.~\eqref{derivativesign} depends on the value of $n$:
\begin{itemize}
    \item[--] If $n>0$, the numerator is positive for $\rho<\rho_\ast e^{1/n}$ and negative for $\rho>\rho_\ast e^{1/n}$. As a consequence, the function $\ln a(\rho)$ reaches its maximum at $\rho=\rho_\ast^{1/n}$, being upper-bounded, so that and condition \eqref{energycondpartial} is  satisfied.
    
    \item[--] If $-1<n\le0$, Eq.~\eqref{derivativesign} and Eq.~\eqref{derivativesignBIS} are both always positive, and a maximum for $\ln a(\rho)$ cannot be found. However, it is easy to show that, when $\rho\gg\rho_\ast$, $\ln a(\rho)$ behaves as $\alpha-\frac{1}{\rho^{1+n}}$, where $\alpha$ is an integration constant. Since $n>-1$, then $\ln a(\rho)$ has an asymptote identified with $\alpha$, and thus it is upper-bounded also in this case.
    
    \item[--] If $n<-1$, one can easily prove that Eq.~\eqref{derivativesign} is always positive. However, in this case, the second term of the denominator dominates over the first and the whole expression behaves as $\rho^{-1}$ for large $\rho$. This means that $\ln a(\rho)$ increases always as $\ln(\rho)$,  so it is not upper-bounded and the condition \eqref{energycondpartial} is never satisfied.
\end{itemize}

\section{Optical properties of logotropic black holes}

In order to study optical properties of the above BH solution, we first focus on the validity of Eq.~\eqref{energycondpartial}, holding for  $\rho>\rho_\ast$, which requires $n>-1$ and thus $\sqrt{a(\rho)}\le1$. Since $n>-1$ ensures that $a(\rho)$ is upper-bounded, we can always find an integration constant for $a(\rho)$ such that the WEC is satisfied. 
Hence, we select this integration constant such that $a(\rho)=1$ at its maximum value. This occurs when $\rho\to\infty$, if $n<0$, and when $\rho=\rho_\ast e^{1/n}$, if $n>0$ (i.e. when $P_\rho=0$). As the first case is unphysical, since it would imply an infinite value of the pressure, we limit our prescription to $n>0$, selecting the function $a(\rho)$ such that $a(\rho_\ast e^{1/n})=1$.
In this way, we ensure that $\sqrt{a(\rho)}\le1$ is satisfied for each $\rho\ne\rho_\ast$ and that $\sqrt{a(\rho)}=1$ when $P$ reaches its maximum.

Therefore, by virtue of Eq.~\eqref{generalmetric}, one easily finds
\begin{equation}
    f(r,\rho)=\sqrt{a(\rho)}-\frac{2M}{r}\,,
    \label{schwsolution}
\end{equation}
implying that the metric \eqref{eq:metric} reads
\begin{equation}\label{pseudoschwmetric}
\begin{split}
    ds^2=\ &\sqrt{a(\rho)}\left(1-\frac{2M}{\sqrt{a(\rho)}r}\right)dt^2\\
    &-\frac{dr^2}{\sqrt{a(\rho)}\left(1-\frac{2M}{\sqrt{a(\rho)}r}\right)}-r^2d\Omega^2\,.
\end{split}
\end{equation}
Adopting the following coordinate transformations
\begin{subequations}
\begin{align}
t&\to (\sqrt{a(\rho)})^{-{1\over2}}t\,,\label{timerescaling}\\
r&\to (\sqrt{a(\rho)})^{{1\over2}}r\,,\label{rrescaling}
\end{align}
\end{subequations}
and preserving the angles, Eq.~\eqref{pseudoschwmetric} describes a Schwarzschild BH with a mass rescaled by
\begin{equation}
    M\rightarrow M(\sqrt{a(\rho)})^{-3/2}\,,
\end{equation}

The stress-energy tensor arising from this metric is 
\begin{equation}\label{pseudoschw stressenergy}
    T^{\mu\nu} =\left(P+\frac{1-\sqrt{a(\rho)}}{8\pi r^2}\right)(e_0^\mu e_0^\nu-e_1^\mu e_1^\nu)-P(e_2^\mu e_2^\nu+e_3^\mu e_3^\nu)\,,
\end{equation}
where the standard Schwarzschild case is recovered as $\sqrt{a(\rho)}\rightarrow1$ and the stress-energy tensor reduces to 
\begin{equation}
    T^{\mu\nu}=\frac{|\Lambda|}{8\pi}\left(e_0^\mu e_0^\nu-\sum_{i=1}^3e_i^\mu e_i^\nu\right)=\frac{|\Lambda|}{8\pi}g^{\mu\nu}\,.
\end{equation}
The latter represents a source for the Einstein field equations that cancels with the term $|\Lambda|g^{\mu\nu}$, thus leading to a vacuum solution and allowing to recover completely the Schwarzschild solution.
In this respect, it appears evident that an Anton-Schmidt fluid is thermodynamically equivalent to a Schwarzschild BH when its density is $\rho=\rho_\ast e^{1/n}$. As the fluid pressure takes its maximum value $P=\frac{A}{n e}$, $P$ being associated to the cosmological pressure, we can fix $A$ as $A=\frac{n e|\Lambda|}{8\pi}$, for $n>0$.

\subsection{The effective refractive index}

We here investigate the properties of our solution arising from the condition $\sqrt{a(\rho)}\ne1$, which implies modifications of the Schwarzschild metric. In particular, we relate this effect to optical properties of the spacetime in presence of a  medium made by logotropic fluids.

To do so, we study the refractive index that could lead to different results due its double interpretation, namely the \textit{optical refractive index} \cite{Perlick:2010zh,Gibbons:2008rj}, $n_o$, and the \textit{Fermat refractive index} \cite{Yi_2011}, $n_F$.
Specifically, we check whether the effects of a logotropic fluid medium are significant to change the optical configuration around a BH. We also show that our procedure is general and can be adapted to other thermodynamic models. 

\subsubsection{Optical refractive index}

The effective optical refractive index, $n_o$, emerges by modifying the metric as \cite{Gibbons:2008rj}
\begin{equation}\label{opticalmetric}
   ds^2= f^2(\hat{r})\left[dt^2-n_o^2\left(d\hat{r}^2+\hat{r}^2d\Omega^2 \right)\right]\,,
\end{equation}
and we can  find plausible transformations to match Eq.~\eqref{pseudoschwmetric} with Eq.~\eqref{opticalmetric}. For the sake of simplicity, we only consider the time rescaling \eqref{timerescaling} neglecting the rescaling of the radius \eqref{rrescaling}. One can easily prove that the refraction indices are the same up to a mass rescaling $M\to M(\sqrt{a(\rho)})^{-1/2}$. Equating the components of the two metrics, we thus obtain
\begin{align}
    f(\hat{r})&=\left(1-\frac{2M}{\sqrt{a(\rho)}r}\right)^{1/2}\,, \label{condtime} 
    \\
    f(\hat{r})n_o(\hat{r})d\hat{r}&=\frac{dr}{\sqrt{\sqrt{a(\rho)}\left(1-\frac{2M}{\sqrt{a(\rho)}r}\right)}}\,, \label{condlong} 
    \\
    f(\hat{r})n_o(\hat{r})\hat{r}&=r\,. \label{condangular}
\end{align}
Dividing Eq.~\eqref{condlong} by Eq.~\eqref{condangular}, we get the following differential equation:
\begin{equation}
    \frac{d\hat{r}}{\hat{r}}=\frac{dr}{r{\sqrt{\sqrt{a}\left(1-\frac{2M}{\sqrt{a}r}\right)}}}\,,
\end{equation}
which, once integrated, gives
\begin{equation}\label{opt solution}
    K\hat{r}^{\sqrt[4]{a}}=r\frac{\sqrt{a}}{M}-1+\sqrt{r\frac{\sqrt{a}}{M}}\sqrt{r\frac{\sqrt{a}}{M}-2}\,,
\end{equation}
where $K$ is an integration constant. Then, inverting Eq.~\eqref{opt solution} yields
\begin{equation}
    r(\hat{r})=\frac{M}{\sqrt{a}}\frac{\left(1+K\hat{r}^{\sqrt[4]{a}}\right)^2}{2K\hat{r}^{\sqrt[4]{a}}}\,.
\end{equation}
so that, from Eq.~\eqref{condtime}, we have
\begin{equation}
    f(\hat{r})=\frac{K\hat{r}^{\sqrt[4]{a}}-1}{K\hat{r}^{\sqrt[4]{a}}+1}\,.
\end{equation}
Hence, the optical refractive index could be computed from Eq.~\eqref{condangular} as
\begin{equation}
    n_o(\hat{r})=\frac{M}{\sqrt{a}}\frac{\left(1+K\hat{r}^{\sqrt[4]{a}}\right)^3}{2K\hat{r}^{\sqrt[4]{a}+1}\left(K\hat{r}^{\sqrt[4]{a}}-1\right)}\,.
\end{equation}
The Schwarzschild case is recovered when $\sqrt{a}=1$. To be consistent with  the notation of \cite{Perlick:2010zh}, in the Schwarzschild case, we set $K=\frac{2}{M}$, thus leading to
\begin{equation}\label{opticalindex}
    n_o(\hat{r})=\frac{1}{\sqrt{a}\hat{r}^{1-\sqrt[4]{a}}}\frac{\left(1+\frac{M}{2\hat{r}^{\sqrt[4]{a}}}\right)^3}{1-\frac{M}{2\hat{r}^{\sqrt[4]{a}}}}.
\end{equation}

Two cases of interest occur for small and large radii. In particular, for small radii, the index of refraction diverges at the horizon radius \cite{Perlick:2010zh,Yi_2011}, i.e. $r_h=\frac{2M}{\sqrt{a}}$, as
\begin{equation}
    n_o(r\to r_h)\sim \frac{2^{\frac{1}{2}+a^{-1/4}}}{M^{a^{-1/4}}}\frac{r_h}{\sqrt{r-r_h}}\,.
\end{equation}

Instead, the case of large radii is not straightforward. In particular,  for $\sqrt{a}\ne1$, the optical refractive index seems to tend to zero, rather than to the unity:
\begin{equation}
    n_o(\hat{r}\to \infty)\sim\frac{1}{\sqrt{a}\hat{r}^{1-\sqrt[4]{a}}}.
\end{equation}
\begin{figure}
    \centering
    \includegraphics[scale=0.67]{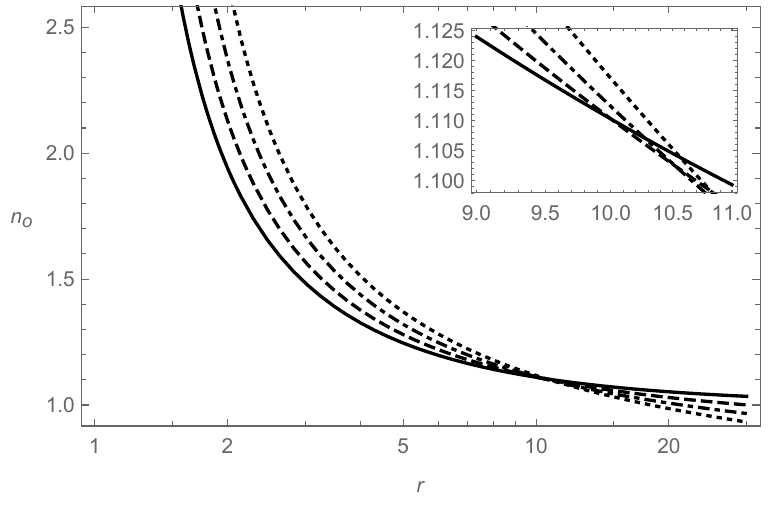}
    \caption{Effective optical refractive index as a function of the external radial coordinate $r$ (distance from the BH center) in a logarithmic scale. The curves correspond to different values of $\sqrt{a(\rho)}$: $\sqrt{a}=1$ (solid), corresponding to the Schwarzschild case); $\sqrt{a}=0.95$ (dashed); $\sqrt{a}=0.9$ (dot-dashed); $\sqrt{a}=0.85$ (dotted). Units of $2M=1$ were adopted.}
    \label{fig: optical index}
\end{figure}
Such a feature is shown in Fig.~\ref{fig: optical index}, where it appears evident that $n_o$ seems to break down at a given $r$, as it does not tend to unity like in the Schwarzschild case. The physical explanation for this behavior is related to the definition of $n_0$. Indeed, at very large radii, the density $\rho$ cannot act as a medium, since the corresponding BH would act as a fully transparent point-like object whose index of refraction cannot depart from $n_o=1$. This can be seen if one does not fix $\sqrt{a(\rho)}$ to a given value, assuming that $\rho$ distributes providing a limiting case for the  Minkowski spacetime. To see that, it is straightforward to notice that, for varying $\rho$, when $\hat r\rightarrow\infty$, $a(\rho)\rightarrow0$ in the denominator of Eq.~\eqref{pseudoschwmetric}, since there is no evidence of logotropic fluids at very large distances. To guarantee asymptotic flatness of Eq.~\eqref{pseudoschwmetric}, one requires $\sqrt{a(\rho)}r\rightarrow\infty $ as both $r\rightarrow\infty$ and $P\rightarrow0$. This can be generalized to any fluid whose density does not explicitly depends on the radial coordinate. Indeed, if one considers $\rho=\rho(\hat r)$, it would be possible to explore the above case without the need to have
$\sqrt{a(\rho)}r\rightarrow\infty $ when both $r\rightarrow\infty$ and $P\rightarrow0$.

The above considerations about the optical refractive index imply that its validity still holds, but it highly depends on the  functional evolution of $\rho$ in terms of $\hat r$. To overcome this issue, one can search for an optical index that, on the contrary, does not take into account how the EoS evolves as a function of the radial coordinate. This subject is investigated in the following, where we deal with effective refractive index resulting from the application of  the Fermat principle.

\subsubsection{Refractive index from the Fermat principle}

As stated above, a simpler approach to investigating the refractive index involves the use of the Fermat principle. This approach turns out to be quite different than $n_0$, because it does not depend on the radial distance $\hat r$.
Its use spans within several gravitational lensing contexts  \cite{Liu:2015wma,Walters:2010gk,Yi_2011}, and it is constructed by considering light rays, i.e. $ds^2=0$, with constant angles. Thus, from Eq.~\eqref{pseudoschwmetric}, we have
\begin{equation}
    dt=\frac{dl}{\sqrt{1-\frac{2M}{\sqrt{a}r}}}\,,
\end{equation}
where $dl^2$ is the spatial part of Eq.~\eqref{pseudoschwmetric}. For a path $\Gamma$, the Fermat principle  reads
\begin{equation}
    \delta\int\frac{d\Gamma}{\sqrt{1-\frac{2M}{\sqrt{a(\rho)}r}}}=0\,.
\end{equation}
Hence, the corresponding refractive index is obtained by comparing the latter with the Fermat principle in the flat case $\delta\int n_F \,d\Gamma=0$:
\begin{equation}\label{Fermat refindex}
    n_F(r)=\left(1-\frac{2M}{\sqrt{a(\rho)}r}\right)^{-{1\over2}}.
\end{equation}
\begin{figure}
    \centering
    \includegraphics[scale=0.67]{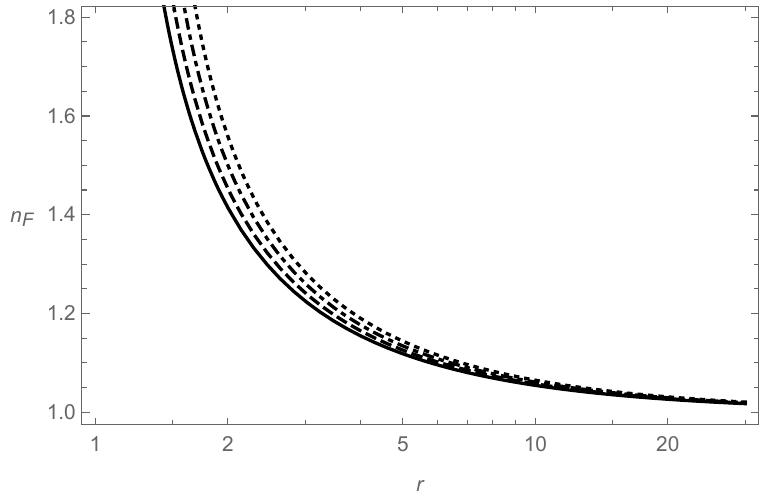}
    \caption{Fermat refractive index as a function of the distance from the BH center $r$ in a logarithmic scale. The different curves correspond to the values of $\sqrt{a(\rho)}$ as in Fig.~\ref{fig: optical index}.}
    \label{refindex}
\end{figure}
\noindent The behavior of the latter, for different values of $\sqrt{a}$, is shown in Fig.~\ref{refindex}. We notice that, for large radii, the Fermat refractive index tends to  $n_F=1$, regardless of the value $\sqrt{a(\rho)}$, consistently  with the fact that, far from the BH, the influence of the fluid cannot be perceived. For small radii, $n_F$ diverges at the event horizon $r_h$ as
\begin{equation}
    n_F(r\to r_h)\sim\sqrt{\frac{r_h}{r-r_h}}.
\end{equation}
\subsubsection{Horizon radius shift}
From both the optical and Fermat refractive indices, it appears evident that the BH event horizon is shifted when $\sqrt{a(\rho)}\ne1$ from $2M$ to $\frac{2M}{\sqrt{a(\rho)}}$. The deviation of the horizon radius from the Schwarzschild case can be expressed as a function of the Anton-Schmidt density as
\begin{equation}\label{horizonradius}
    \Delta r_h(\rho)=\frac{r_h(\rho)-2M}{2M}=\frac{1}{\sqrt{a(\rho)}}-1\,.
\end{equation}
\begin{figure}
    \centering
    \includegraphics[scale=0.67]{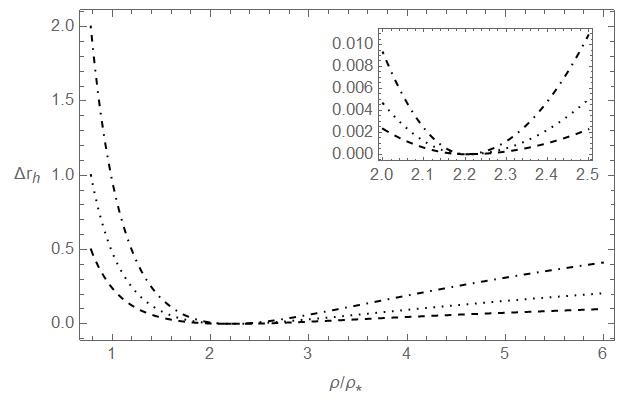}
    \caption{Deviation of the horizon radius as a function of the density of the Anton-Schmidt fluid. The values on the $x$ and $y$ axes are expressed in powers of $10^{4}$ and $10^{-6}$, respectively. The parameter $n$ is fixed to $0.1$, while the different curves correspond to different values of $\rho_\ast$: $\rho_\ast=0.5 A$ (solid), $\rho_\ast=A$ (dotted) and $\rho_\ast=2A$ (dashed). The Schwarzschild solution is recovered at the minima of the curves, i.e. when $\rho/\rho_\star=e^{1/n}\simeq 22\cdot10^3.$ The top-right subplot enhances the behavior of $r_h-1$ around such minimum.}
    \label{massscaling}
\end{figure}
A plot of the latter quantity is provided in Fig.~\ref{massscaling}. It is interesting to notice how different values of the Anton-Schmidt parameters (namely, different ratios $\rho_\ast/A$) affect the behavior of the horizon radius in the optical framework. Namely, the greater $\rho_\ast/A$, the smaller deviations of the horizon radius from the Schwarzschild one occur as $\rho$ deviates from $\rho_\ast e^{1/n}$.

\section{Outlook and perspectives}
\label{sec:conclusions}

In this paper, we investigated a class of asymptotic AdS BH metrics whose thermodynamics matches that of logotropic models. For this purpose, we considered the Anton-Schmidt fluid, which contains the pure logotropic case as a limiting case. We thus developed a general method to obtain a metric solution and, by requiring the presence of a BH horizon, we found plausible metric functions describing the physical scenario under study. 
Our treatment has been carried out solving the Einstein field equations, i.e., requiring our solution to be physical BHs.

We then studied the most suitable values of the free parameters associated with our class of solutions. 
In particular, taking into account the  energy conditions for the source fluid, we showed that a BH solution of the  Einstein field equations is obtained for an asymptotic AdS spacetime, guaranteeing that the Einstein equations fully hold in presence of the cosmological constant, and leading to two specific cases: i) a pure constant pressure, corresponding to a constant density of the Anton-Schmidt fluid; ii) an exotic outcome, which should be physically discarded. In this respect, we checked whether, by relaxing the SEC, more general classes of solutions may be found. In particular, our analysis demonstrates that an Anton-Schmidt fluid with positive pressure is capable of emulating the thermodynamics of a spherically-symmetric compact object as soon as the pressure of the fluid is maximized. 

Furthermore, we analyzed the optical properties of our solutions. To do so, we followed the standard recipe to obtain an effective refractive index in spherical coordinates. We thus adopted two definitions, namely the optical and the Fermat refractive indexes. In the first case, we showed the limits of the model predictions, which mostly require knowing how the density evolves as a function of the radial distance, and provides an unphysical behavior for large radii. Moreover, to overcome the aforementioned issues, we considered the Fermat refractive index. 
In so doing, we discussed the functional dependence on $r$, with particular regard to its asymptotic values. Likely divergences at small radii were discussed, whereas a direct comparison with the Schwarzschild case was prompted, showing where the deviations from the latter case are much more evident. Quite clearly, the corresponding optical effects imply that the refractive indexes increase due to the presence of the thermodynamic medium constituted by the Anton-Schmidt fluids. 

Future works will focus on alternative fluids characterized by thermodynamic effects that are mainly different from logotropic models. In particular, it would be interesting to take into account real fluids and show, for instance, how transition phases could affect these scenarios. Another plausible extension would be to promote our treatment to more general compact objects \cite{Cardoso:2019rvt} and/or on accretion disk contexts \cite{DAgostino:2022ckg,Kurmanov:2021uqv,Boshkayev:2021chc}. A crucial point deserving future efforts will be investigating the relation between the thermodynamic properties of the fluid and the BH, in order to enable a richer use of thermodynamics and make the extended phase approach more robust and fully justified from a physical perspective.

\begin{acknowledgements}
The authors are grateful to the anonymous referee for suggestions that helped to improve the quality of the manuscript. S.C. and R.D. acknowledge the support of Istituto Nazionale di Fisica Nucleare (INFN), \textit{iniziative specifiche} MOONLIGHT2 and QGSKY. A.L. and O.L. would like to thank Roberto Giambò and Stefano Mancini for helpful discussions. O.L. acknowledges the Ministry of Education and Science of the Republic of Kazakhstan, Grant: IRN AP08052311. 
\end{acknowledgements}

\appendix

\section{Suitable forms of $h(r,\rho)$} 
\label{appendixA}
As mentioned in Sec.~\ref{BHth}, the ansatz \eqref{addLinComb} considered in \cite{Rajagopal:2014ewa,Setare:2015xaa,Debnath:2019mzs} made use of specific functions for $R_i(\rho)$. However, since we do not know $\emph{a priori}$ the best functions to choose, we describe here a general procedure aimed at minimizing the loss of generality of the solution.\\

Resuming, by inserting the solution \eqref{addLinComb} (once the functions $R_i(\rho)$ are known) into Eq.~\eqref{diffeq}, the latter takes the form of Eq.~\eqref{eqs}. By solving the equations $\{\xi_j(r)F_j(\rho)=0\}_j$ separately, we get the functions $X_i(r)$ from the equations $\{\xi_j(r)=0\}_j$. The equations $\xi_j(r)=0$ for $X_i(r)$ can be algebraic or differential. In particular, for a given $j$, i.e. considering $\xi_j(r)F_j(\rho)=0$, we can distinguish two cases:
\begin{itemize}
    \item[--] if the function $F_j(\rho)$ is present in the third term of the left-hand side of Eq.~\eqref{diffeq}, then the equation $\xi(r)=0$ is differential due to the  presence of a partial derivative with respect to $r$;
    \item[--] if the function $F_j(\rho)$ is not present in the third term of the left-hand side of Eq.~\eqref{diffeq}, but it is in some of the other two terms, then the equation $\xi(r)$ is algebraic.
\end{itemize}

Since we search for a solution as general as possible, we want all the equations for $X_i(r)$, i.e. $\{\xi_j(r)=0\}_j$ to be differential, so that for each one of them we can get an integration constant. For this purpose, we choose the functions $R_i(\rho)$ such that all the independent functions of $\rho$, i.e. $F_j(\rho)$, present in the first and second term, must be present also in the third term.
To do so, we consider the independent functions of $\rho$ present in each term. Therefore, we have
\begin{subequations}\label{systemcomb}
\begin{align}
     &8\pi r^2(P-2\rho)P_\rho=\mathfrak{L}(PP_\rho,\rho P_\rho)\,,\\
     &6(\rho+P)h_\rho=\mathfrak{L}(\rho (R_i(\rho))_\rho,P (R_i(\rho))_\rho)\,,\\
     &-3(rh)'P_\rho=\mathfrak{L}(P_\rho R_i(\rho))\,,
\end{align}
\end{subequations}
where $\mathfrak{L}(\dots)$ means ``linear combination of $\dots$".
First, we set $R_1(\rho)=1$ (the motivation will be clarified later). Then, to have the functions of $\rho$ in the first term included in the third one, we take $R_2(\rho)=\rho$ and $R_3(\rho)=P(\rho)$. We thus have $h=\mathfrak{L}(1,\rho,P)\Longrightarrow h_\rho=\mathfrak{L}(1,P_\rho)$. Then, the third term becomes $\mathfrak{L}(P_\rho,\rho P_\rho, PP_\rho)$ and the second one $\mathfrak{L}(\rho,P,\rho P_\rho, PP_\rho)$. At this stage, we should take into account the specific fluid under study in order to check whether some of the terms of  above linear combinations are linearly dependent. 

In particular, recalling Anton-Schmidt's pressure \eqref{ASfluid}, we have
\begin{equation}
    P\equiv \mathfrak{L}(P_1(\rho),P_2(\rho))\,,
\end{equation}
where $P_1(\rho)= \left(\frac{\rho}{\rho_\ast}\right)^{-n}\ln\rho$ and $P_2(\rho)=\left(\frac{\rho}{\rho_\ast}\right)^{-n}$. Taking the derivative of $P$ with respect to the density, we obtain
\begin{align}
    P_\rho=\frac{1}{\rho}\left[-nA(P_1(\rho)-\ln\rho_\ast P_2(\rho))+AP_2(\rho)\right].
\end{align}
Hence, $\rho P_\rho=\mathfrak{L}(P_1,P_2)$, so that both $\rho P_\rho$ and $P$ are different linear combinations of the same functions $P_1$ and $P_2$. The third term becomes $\mathfrak{L}(P_\rho,P_1,P_2,PP_\rho)$, while the second one reads $\mathfrak{L}(\rho, P_1, P_2, PP_\rho)$. As not all the terms in the second term are present in the third, we can add to $h$ a term proportional to $b(\rho)$, so that $h=\mathfrak{L}(1,\rho,P,b(\rho))$. In this way, the third term becomes $\mathfrak{L}(P_\rho,P_1,P_2,PP_\rho,P_\rho b(\rho))$ and the second $\mathfrak{L}(\rho, P_1, P_2, PP_\rho, \rho b_\rho(\rho),Pb_\rho(\rho))$. 
To have all the functions of $\rho$ in the second term and also in the third one, we need the function $b(\rho)$ to be a linear combination of $\rho$, $\rho  b_\rho$ and $Pb_\rho$, namely
\begin{equation}\label{effediffeq}
    P_\rho b(\rho)=B_1\rho+B_2 \rho b_\rho+B_3 P b_\rho\,,
\end{equation}
where $B_1$, $B_2$ and $B_3$ are generic constants. Finally, the ansatz for the function $h(\rho, r)$ can be written as
\begin{equation}\label{lessgeneralityresultansatz}
    h(r,\rho)=X_1(r)+X_2(r)\rho+X_3(r)P_1+X_4(r) P_2+X_5(r)b(\rho)\,,
\end{equation}
with $b(\rho)$ satisfying Eq.~\eqref{effediffeq}.\\

\section{Derivation of the metric function}\label{appendixB}

In order to derive the unknown function $h(r,\rho)$, we report below some useful relations:
\begin{subequations}
\begin{align}
    &P=AP_1(\rho)-A\ln\rho_\ast P_2(\rho)\,, \label{PrP1P2} \\
    &\rho P_\rho=-nA(P_1(\rho)-\ln\rho_\ast P_2(\rho))+AP_2(\rho)\,, \\
    &\rho P_{1,\rho}=-nP_1+P_2\,, \\
    & \rho P_{2,\rho}=-nP_2\,.
\end{align}
\end{subequations}
Thus, inserting Eq.~\eqref{lessgeneralityresultansatz} into Eq.~\eqref{diffeq}, we obtain 
\begin{align}\label{macrodiffeq}
        &8\pi r^2\left(A^2P_1P_{1,\rho}-A^2\ln\rho_\ast P_1P_{2,\rho}-A^2\ln\rho_\ast P_2P_{1,\rho}\right. \nonumber \\ 
        &\left.+A^2\ln^2\rho_\ast  P_2P_{2,\rho}-2A(1+n\ln\rho_\ast)P_2+2nAP_1\right) \nonumber \\
        &+6\left(X_2\rho-nX_3P_1+(X_3-nX_4)P_2+X_5\rho b_\rho+AP_1X_2\right. \nonumber \\
        &+AX_3P_1P_{1,\rho}+AX_4P_1P_{2,\rho}+AX_5P_1b_\rho-A\ln\rho_\ast X_2P_2 \nonumber \\
        &\left.-A\ln\rho_\ast X_3P_2P_{1,\rho}-A\ln\rho_\ast X_4P_2P_{2,\rho}-A\ln\rho_\ast X_5 P_2b_\rho\right)\nonumber\\
        &-3\left(A(rX_1)'P_{1,\rho}-nA(rX_2)'P_1+A(1+n\ln\rho_\ast)(rX_2)'P_2\right. \nonumber\\
        &+A(rX_3)'P_1P_{1,\rho}+A(rX_4)'P_2P_{1,\rho}-A\ln\rho_\ast(rX_1)'P_{2,\rho} \nonumber\\
        &-A\ln\rho_\ast(rX_3)'P_1P_{2,\rho}-A\ln\rho_\ast(rX_4)'P_2P_{2,\rho}+B_1(rX_5)'\rho \nonumber\\
        &\left.+B_2(rX_5)'\rho b_\rho+B_3(rX_5)'Pb_\rho\right)=0\,,
\end{align}
where the last three terms have been obtained by making use of Eq.~\eqref{effediffeq}.
The independent functions of $\rho$ are $\rho$, $P_1$, $P_2$, $P_{1,\rho}$, $P_{2,\rho}$, $P_1P_{1,\rho}$, $P_2P_{2,\rho}$, $P_1P_{2,\rho}$, $P_2P_{1,\rho}$, $\rho b_\rho$, $P b_\rho$. Hence, we have $11$ differential equations for the functions $\{X_i(r)\}_{i=1}^5$. Namely, the differential equation $\xi_j(r)$ corresponding to each $F_j(\rho)$ are as follows:
\begin{subequations}
\begin{align}
   \rho:\quad & 6X_2-3B_1(rX_5)'=0\,; \label{rho}\\
   P_1:\quad & 16\pi nAr^2+6(AX_2-nX_3)+3nA(rX_2)'=0\,; \label{P1}\\
   P_2:\quad &-16\pi r^2A(1+n\ln\rho_\ast)+6(X_3-nX_4)\label{P2}\\
   &-6A\ln\rho_\ast X_2-3A(1+n\ln\rho_\ast)(rX_2)'=0\,; \nonumber \\
   P_{1,\rho}:\quad & 3A(rX_1)'=0\,;\label{easy1}\\
   P_{2,\rho}:\quad & 3A(rX_1)'\ln\rho_\ast=0\,;\label{easy2}\\
   P_1P_{1,\rho}:\quad & 8\pi r^2A^2+6AX_3-3A(rX_3)'=0\,;\label{noPressure1}\\
   P_1P_{2,\rho}:\quad &8\pi r^2A^2\ln\rho_\ast-6AX_4-3A\ln\rho_\ast(rX_3)'=0\,; \label{noPressure2} \\
   P_2P_{1,\rho}:\quad &8\pi r^2A^2\ln\rho_\ast+6A\ln\rho_\ast X_3+3A(rX_4)'=0\,;\label{noPressure3}\\
   P_2P_{2,\rho}:\quad & 8\pi r^2A^2\ln^2\rho_\ast-6A\ln\rho_\ast X_4+3A\ln\rho_\ast(rX_4)' \nonumber \label{noPressure4}\\
   &=0\,;\\
   \rho b_\rho:\quad & 6X_5-3B_2(rX_5)'=0\,;\label{medium1}\\
   P b_\rho:\quad & 6X_5-3B_3(rX_5)'=0\,.\label{medium2}
\end{align}
\end{subequations}
We can immediately notice that Eqs.~\eqref{easy1} and \eqref{easy2} are degenerate, providing 
\begin{equation}\label{X1}
    X_1(r)=\frac{c_1}{r}\,,
\end{equation}
where $c_1$ is a constant. From Eqs.~\eqref{medium1} and \eqref{medium2}, we thus obtain
\begin{align}
    B_2&=B_3\,,  \label{coeff1} \\
    X_5(r)&=c_2r^{\frac{2}{B_2}-1}, \label{X5}
\end{align}
where $c_2$ is an integration constant. In virtue of \eqref{X5}, we can easily compute Eq.~\eqref{rho} obtaining
\begin{equation}\label{X2}
    X_2(r)=\frac{B_1}{B_2}X_5(r)=c_1\frac{B_1}{B_2}r^{\frac{2}{B_2}-1}.
\end{equation}
Then, Eq.~\eqref{noPressure1} can be written as
\begin{equation}
    X_3'-\frac{X_3}{r}=\frac{8}{3}\pi Ar\,,
\end{equation}
whose solution is
\begin{align}\label{X3}
    X_3(r)&=\exp{\left(\int\frac{dr}{r}\right)}\left(\int\frac{8}{3}\pi A r\exp{\left(\int-\frac{dr}{r}\right)}+c_3\right) \nonumber \\
    &=c_3r+\frac{8}{3}\pi r^2A\,,
\end{align}
where $c_3$ is a further integration constant. The function $X_4(r)$ can be obtained from Eq.~\eqref{noPressure4} as
\begin{equation}\label{X4}
    X_4(r)=c_4r-\frac{8}{3}\pi r^2A\ln\rho_\ast\,,
\end{equation}
with $c_4$ being an integration constant. Inserting Eqs.~\eqref{X3} and \eqref{X4} into Eqs.~\eqref{noPressure2} and \eqref{noPressure3}, we obtain the following relation between the integration constants $c_3$ and $c_4$:
\begin{equation}\label{coeff2}
    c_4=-c_3\ln\rho_\ast\,.
\end{equation}
From this, we find
\begin{equation}
    X_4(r)=-\ln\rho_\ast X_3(r)\,.
\end{equation}
Using the results obtained for the functions $X_i(r)$ and the restrictions on the coefficients given by Eqs.~\eqref{coeff1} and \eqref{coeff2}, from Eqs.~\eqref{P1} and \eqref{P2} we find, respectively,
\begin{align}
   &nc_3r-Ac_2\frac{B_1}{B_2}\left(1+\frac{n}{B_2}\right)r^{\frac{2}{B_2}-1}=0\,, \label{finS1} \\
    & (1+n\ln\rho_\ast)c_3r-Ac_2\frac{B_1}{B_2}\left(\ln\rho_\ast+\frac{1+n\ln\rho_\ast}{B_2}\right)r^{\frac{2}{B_2}-1}=0\,. \label{finS2}
\end{align}
From the study of Eqs.~\eqref{finS1} and \eqref{finS2}, we can infer the free coefficients $c_2$, $c_3$, $B_1$ and $B_2$ and, thus, determine $h(\rho, r)$. The various possibilities are listed below.
\begin{itemize}
    \item[--] The simplest case is when $c_3=c_2=0$, for which one has
    \begin{equation}
        \begin{cases}
        &X_1(r)=\dfrac{c_1}{r}\,,\\
        &X_2(r)=X_5(r)=0\,,\\
        &X_3(r)=\dfrac{8}{3}\pi r^2A\,,\\
        &X_4(r)=-\dfrac{8}{3}\pi r^2 A\ln\rho_\ast \,.
        \end{cases}
    \end{equation}
    In this case, we obtain
    \begin{align}\label{addtermfake}
        h(r,\rho)&=\frac{c_1}{r}+\frac{8}{3}\pi r^2(AP_1(\rho)-A\ln\rho_\ast P_2(\rho)) \nonumber \\ 
        &=\frac{c_1}{r}+\frac{8}{3}\pi r^2 P(\rho)\,,
    \end{align}
    where the last equality is due to \eqref{PrP1P2}.
    \item[--] Suppose $c_2$, $c_3$, $B_1\ne0$. In this case, the system made by Eqs.~\eqref{finS1} and \eqref{finS2} becomes
    \begin{equation}\label{systemsol}
        \begin{cases}
        &\dfrac{B_2c_3}{Ac_2B_1}r=\left(\dfrac{1+\frac{n}{B_2}}{n}\right)r^{\frac{2}{B_2}-1}\,,\\
        &\dfrac{B_2c_3}{Ac_2B_1}r=\left(\dfrac{\ln\rho_\ast+\frac{1+n\ln\rho_\ast}{B_2}}{1+n\ln\rho_\ast}\right)r^{\frac{2}{B_2}-1}\,.
        \end{cases}
    \end{equation}
    Since the system must be valid for each $r$, we need $B_2=1$ so that the exponents on $r$ in both sides are the same. Then, we have
    \begin{equation}
        \frac{1+n}{n}=\frac{\ln\rho_\ast+1+n\ln\rho_\ast}{1+n\ln\rho_\ast}\,,
    \end{equation}
    or, equivalently,
    \begin{equation}
        1+n+n\ln\rho_\ast+n^2\ln\rho_\ast=n+\ln\rho_\ast+n^2\ln\rho_\ast\,,
    \end{equation}
   which admits no solutions.
   
    \item[--] Suppose $c_3=0$, but $B_1\neq 0$ and $c_2\neq 0$. In this case, the left hand sides of system \eqref{systemsol} become
    \begin{equation}
    \begin{cases}
        &1+\dfrac{n}{B_2}=0\,,\\
        &\ln\rho_\ast+\dfrac{1+n\ln\rho_\ast}{B_2}=0\,.
    \end{cases}
    \end{equation}
    From the first equation, we obtain $B_2=-n$, so that 
    \begin{equation}
        \ln\rho_\ast-\frac{1}{n}-\ln\rho_\ast=0\Longrightarrow \frac{1}{n}=0\,.
    \end{equation}
    This is possible only in the limit $n\rightarrow\infty$, corresponding to a pressureless fluid.
    
    \item[-] Consider $c_3=B_1=0$ and $c_2\ne0$. The functions $X_i(r)$ then read
    \begin{equation}
        \begin{cases}
        &X_1(r)=\dfrac{c_1}{r}\,,\\
        &X_2(r)=0\,,\\
        &X_3(r)=\dfrac{8}{3}\pi r^2A\,, \\
        &X_4(r)=-\dfrac{8}{3}\pi r^2 A\ln\rho_\ast\,, \\
        &X_5(r)=c_2r^{\frac{2}{B_2}-1}\,.
        \end{cases}
    \end{equation}
    Therefore, we finally obtain
    \begin{equation}
        h(r,\rho)=\frac{c_1}{r}+\frac{8}{3}\pi r^2P(\rho)+c_2r^{\frac{2}{B_2}-1}(a(\rho))^{1/B_2}\,.
    \end{equation}
    
\end{itemize}


\begin{thebibliography}{99}

\bibitem{Hawking:1971vc}
S.~W.~Hawking,  Commun. Math. Phys. \textbf{25}, 152 (1972).

\bibitem{Bardeen:1973gs}
J.~M.~Bardeen, B.~Carter and S.~W.~Hawking, Commun. Math. Phys. \textbf{31}, 161 (1973).

\bibitem{Bekenstein:1973ur}
J.~D.~Bekenstein, Phys. Rev. D \textbf{7}, 2333 (1973).

\bibitem{Hawking:1975vcx}
S.~W.~Hawking, Commun. Math. Phys. \textbf{43}, 199 (1975) [Erratum: Commun. Math. Phys. \textbf{46}, 206 (1976)].

\bibitem{Wald:1999vt}
R.~M.~Wald, Living Rev. Rel. \textbf{4}, 6 (2001).

\bibitem{Good:2020qsy}
M.~R.~R.~Good and Y.~C.~Ong, Eur. Phys. J. C \textbf{80}, 1169 (2020).

\bibitem{Myung:2007av}
Y.~S.~Myung, Y.~W.~Kim and Y.~J.~Park, Gen. Rel. Grav. \textbf{41}, 1051 (2009).

\bibitem{Ghaffarnejad:2013wpa}
H.~Ghaffarnejad, Astrophys. Space Sci. \textbf{361}, 7 (2016).

\bibitem{JahaniPoshteh:2021clv}
M.~B.~Jahani Poshteh and R.~B.~Mann, Phys. Rev. D \textbf{103}, 104024 (2021).

\bibitem{Cai:2009qs}
R.~G.~Cai, L.~M.~Cao and N.~Ohta, Phys. Lett. B \textbf{679}, 504 (2009).


\bibitem{Liu:2016urf}
H.~Liu and X.~h.~Meng, Mod. Phys. Lett. A \textbf{32}, 1750146 (2017).

\bibitem{Gibbons:1977mu}
G.~W.~Gibbons and S.~W.~Hawking, Phys. Rev. D \textbf{15}, 2738 (1977).

\bibitem{Pappas:2017kam}
T.~Pappas and P.~Kanti, Phys. Lett. B \textbf{775}, 140 (2017).

\bibitem{Davies:1988dma}
P.~C.~W.~Davies, Ann. Inst. H. Poincare Phys. Theor. \textbf{49}, 297 (1988).

\bibitem{Chatterjee:2020gfc}
B.~Chatterjee and N.~Banerjee, Eur. Phys. J. C \textbf{81}, 604 (2021). 

\bibitem{Donnay:2019zif}
L.~Donnay and G.~Giribet, Class. Quant. Grav. \textbf{36}, 165005 (2019).

\bibitem{Anabalon:2012tu}
A.~Anabalon and A.~Cisterna, Phys. Rev. D \textbf{85}, 084035 (2012).

\bibitem{Hawking:1982dh}
S.~W.~Hawking and D.~N.~Page, Commun. Math. Phys. \textbf{87}, 577 (1983).

\bibitem{Maldacena:1997re}
J.~M.~Maldacena, Adv. Theor. Math. Phys. \textbf{2}, 231 (1998).

\bibitem{Lunin:2001jy}
O.~Lunin and S.~D.~Mathur, Nucl. Phys. B \textbf{623}, 342 (2002).

\bibitem{Hubeny:2014bla}
V.~E.~Hubeny, Class. Quant. Grav. \textbf{32}, 124010 (2015).

\bibitem{Rajagopal:2014ewa}
A.~Rajagopal, D.~Kubiz\v{n}\'ak and R.~B.~Mann, Phys. Lett. B \textbf{737}, 277 (2014).

\bibitem{Roy:2021ucl}
T.~Roy and U.~Debnath, Int. J. Mod. Phys. A  \textbf{36}, 2150114 (2021).

\bibitem{Capozziello:2004ej}
S.~Capozziello, V.~F.~Cardone, S.~Carloni, S.~De Martino, M.~Falanga, A.~Troisi and M.~Bruni,  JCAP \textbf{04}, 005 (2005).

\bibitem{Setare:2015xaa}
M.~R.~Setare and H.~Adami, Phys. Rev. D \textbf{91}, 084014 (2015).

\bibitem{Debnath:2019mzs}
U.~Debnath, Eur. Phys. J. Plus \textbf{135}, 424 (2020).

\bibitem{Kamenshchik:2001cp}
A.~Y.~Kamenshchik, U.~Moschella and V.~Pasquier, Phys. Lett. B \textbf{511}, 265 (2001).

\bibitem{Chavanis:2015paa}
P.~H.~Chavanis, Eur. Phys. J. Plus \textbf{130}, 130 (2015).

\bibitem{Chavanis:2015eka}
P.~H.~Chavanis, Phys. Lett. B \textbf{758}, 59 (2016).

\bibitem{Chavanis:2016pcp}
P.~H.~Chavanis and S.~Kumar, JCAP \textbf{05}, 018 (2017).

\bibitem{Benaoum:2021pqf}
H.~B.~Benaoum, P.~H.~Chavanis and H.~Quevedo, arXiv:2112.13318 [gr-qc].

\bibitem{ANTON1997449}
H. Anton and P. C. Schmidt, Intermetallics \textbf{5}, 449 (1997).

\bibitem{ASfluid}
B. Mayer, H. Anton, E. Bott, M. Methfessel, J. Sticht, J. Harris and P. C. Schmidt, Intermetallics \textbf{11}, 23 (2003).

\bibitem{Capozziello:2017buj}
S.~Capozziello, R.~D'Agostino and O.~Luongo, Phys. Dark Univ. \textbf{20}, 1 (2018).

\bibitem{Capozziello:2018mds}
S.~Capozziello, R.~D'Agostino, R.~Giamb\`o and O.~Luongo, Phys. Rev. D \textbf{99}, 023532 (2019).

\bibitem{Boshkayev:2019qcx}
K.~Boshkayev, R.~D'Agostino and O.~Luongo, Eur. Phys. J. C \textbf{79}, 332 (2019).

\bibitem{Odintsov:2018obx}
S.~D.~Odintsov, V.~K.~Oikonomou, A.~V.~Timoshkin, E.~N.~Saridakis and R.~Myrzakulov,  Annals Phys. \textbf{398}, 238 (2018).

\bibitem{DAgostino:2019wko}
R.~D'Agostino, Phys. Rev. D \textbf{99}, 103524 (2019).

\bibitem{DAgostino:2021vvv}
R.~D'Agostino and O.~Luongo, Phys. Lett. B \textbf{829}, 137070 (2022).

\bibitem{Capozziello:2019cav}
S.~Capozziello, R.~D'Agostino and O.~Luongo, Int. J. Mod. Phys. D \textbf{28}, 1930016 (2019).

\bibitem{Capozziello:2022wgl}
S.~Capozziello and R.~D'Agostino, Phys. Lett. B \textbf{832}, 137229 (2022).

\bibitem{Capozziello:2022rac}
S.~Capozziello, R.~D'Agostino and O.~Luongo, Phys. Lett. B \textbf{834}, 137475 (2022).

\bibitem{Bajardi:2022tzn}
F.~Bajardi and R.~D'Agostino, arXiv:2208.02677 [gr-qc].

\bibitem{DAgostino:2022tdk}
R.~D'Agostino and R.~C.~Nunes, Phys. Rev. D \textbf{106}, 124053 (2022).


\bibitem{Tolman:1939jz}
R.~C.~Tolman, Phys. Rev. \textbf{55}, 364 (1939).

\bibitem{Oppenheimer:1939ne}
J.~R.~Oppenheimer and G.~M.~Volkoff, Phys. Rev. \textbf{55}, 374 (1939).

\bibitem{Capozziello:2020szy}
S.~Capozziello, R.~D'Agostino and D.~Gregoris, Phys. Dark Univ. \textbf{28}, 100513 (2020).

\bibitem{Gunasekaran:2012dq}
S.~Gunasekaran, R.~B.~Mann and D.~Kubiznak, JHEP \textbf{11}, 110 (2012).

\bibitem{Kubiznak:2012wp}
D.~Kubiznak and R.~B.~Mann, JHEP \textbf{07}, 033 (2012).

\bibitem{Chavanis:2022vzi}
P.~H.~Chavanis, Phys. Dark Univ. \textbf{37}, 101098 (2022).

\bibitem{Boshkayev:2021uvk}
K.~Boshkayev, T.~Konysbayev, O.~Luongo, M.~Muccino and F.~Pace, Phys. Rev. D \textbf{104}, 023520 (2021).

\bibitem{Luongo:2018lgy}
O.~Luongo and M.~Muccino, Phys. Rev. D \textbf{98}, 103520 (2018).

\bibitem{Belfiglio:2022egm}
A.~Belfiglio, R.~Giamb\`o and O.~Luongo, arXiv:2206.14158 [gr-qc].

\bibitem{DAgostino:2022fcx}
R.~D'Agostino, O.~Luongo and M.~Muccino, Class. Quant. Grav. \textbf{39}, 195014 (2022).

\bibitem{Bekenstein:1974ax}
J.~D.~Bekenstein, Phys. Rev. D \textbf{9}, 3292 (1974).

\bibitem{Bekenstein:1975tw}
J.~D.~Bekenstein, Phys. Rev. D \textbf{12}, 3077 (1975).


\bibitem{Kastor:2009wy}
D.~Kastor, S.~Ray and J.~Traschen, Class. Quant. Grav. \textbf{26}, 195011 (2009).

\bibitem{Dolan:2010ha}
B.~P.~Dolan, Class. Quant. Grav. \textbf{28}, 125020 (2011).


\bibitem{Luongo:2014qoa}
O.~Luongo and H.~Quevedo, Phys. Rev. D \textbf{90}, 084032 (2014).

\bibitem{Luongo:2015zaa}
O.~Luongo and H.~Quevedo, Found. Phys. \textbf{48}, 17 (2018).

\bibitem{Giambo:2020jjo}
R.~Giamb\`o, O.~Luongo and H.~Quevedo, Phys. Dark Univ. \textbf{30}, 100721 (2020).

\bibitem{Papapetrou:1966zz}
A.~Papapetrou, Ann. Inst. H. Poincare Phys. Theor. \textbf{4}, 83 (1966).

\bibitem{Stephani:2003tm}
H.~Stephani, D.~Kramer, M.~A.~H.~MacCallum, C.~Hoenselaers and E.~Herlt, Cambridge Univ. Press (2003).



\bibitem{mima}
J. P. S. Lemos, O. B. Zaslavskii, Phys. Rev. D \textbf{78}, 024040 (2008).

\bibitem{kis}
V. V. Kiselev,  Class. Quant. Grav. \textbf{20}, 1187 (2003).


\bibitem{vis}
M. Visser, Class. Quant. Grav. \textbf{37},  4, 045001 (2020).

\bibitem{Bayin86}
S. S. Bayin, Astrop. J. \textbf{303}, 101 (1986).




\bibitem{Gibbons:2008rj}
G.~W.~Gibbons and M.~C.~Werner, Class. Quant. Grav. \textbf{25}, 235009 (2008).

\bibitem{Perlick:2010zh}
V.~Perlick, arXiv:1010.3416 [gr-qc].

\bibitem{Yi_2011}
Y. G. Yi, Astrophys. Sp. Sci. \textbf{336}, 437 (2011).

\bibitem{Walters:2010gk}
S.~J.~Walters, L.~K.~Forbes and P.~D.~Jarvis, Mon. Not. Roy. Astron. Soc. \textbf{409}, 953 (2010).

\bibitem{Liu:2015wma}
H.~Liu, X.~Wang, H.~Li and Y.~Ma, Eur. Phys. J. C \textbf{77}, 723 (2017).

\bibitem{Cardoso:2019rvt}
V.~Cardoso and P.~Pani, Living Rev. Rel. \textbf{22}, 4 (2019).

\bibitem{DAgostino:2022ckg}
R.~D'Agostino, R.~Giamb\`o and O.~Luongo, arXiv:2204.02098 [gr-qc].

\bibitem{Kurmanov:2021uqv}
E.~Kurmanov, K.~Boshkayev, R.~Giamb\`o, T.~Konysbayev, O.~Luongo, D.~Malafarina and H.~Quevedo, Astrophys. J. \textbf{925}, 210 (2022).

\bibitem{Boshkayev:2021chc}
K.~Boshkayev, T.~Konysbayev, E.~Kurmanov, O.~Luongo, D.~Malafarina and H.~Quevedo, Phys. Rev. D \textbf{104}, 084009 (2021).

\end{thebibliography}
\end{document}